\documentclass[10pt,aps,prl,twocolumn,groupedaddress,showkeys]{revtex4-1}

\usepackage{graphicx}  
\usepackage{dcolumn}   
\usepackage{bm}        
\usepackage{tabularx}
\usepackage{amsmath}
\usepackage{color,soul}  

\begin{document}

\title{First principles study of charge diffusion between proximate solid state qubits and its implications on sensor applications}

\author{Jyh-Pin Chou}
\affiliation{Institute for Solid State Physics and Optics, 
 Wigner Research Centre for Physics, Hungarian Academy of Sciences, 
 Budapest, POB 49, H-1525, Hungary}

\author{Zolt\'{a}n Bodrog}
\affiliation{Institute for Solid State Physics and Optics, 
 Wigner Research Centre for Physics, Hungarian Academy of Sciences, 
 Budapest, POB 49, H-1525, Hungary}

\author{Adam Gali} \email{gali.adam@wigner.mta.hu}
\affiliation{Institute for Solid State Physics and Optics, 
 Wigner Research Centre for Physics, Hungarian Academy of Sciences, 
 Budapest, POB 49, H-1525, Hungary}
\affiliation{Department of Atomic Physics, 
 Budapest University of Technology and Economics, 
 Budafoki \'ut 8, H-1111, Budapest, Hungary}

\date{\today}

\begin{abstract}
Solid state qubits from paramagnetic point defects in solids are promising platforms to realize quantum networks and novel nanoscale sensors. Recent advances in materials engineering make possible to create proximate qubits in solids that might interact with each other, leading to electron spin/charge fluctuation. Here we develop a method to calculate the tunneling-mediated charge diffusion between point defects from first principles, and apply it to nitrogen-vacancy (NV) qubits in diamond. The calculated tunneling rates are in quantitative agreement with previous experimental data. Our results suggest that proximate neutral and negatively charged NV defect pairs can form an NV--NV molecule. A tunneling-mediated model for the source of decoherence of the near-surface NV qubits is developed based on our findings on the interacting qubits in diamond.
 \end{abstract}


\maketitle

\begin{figure}[ht]
\includegraphics[width=0.48\textwidth]{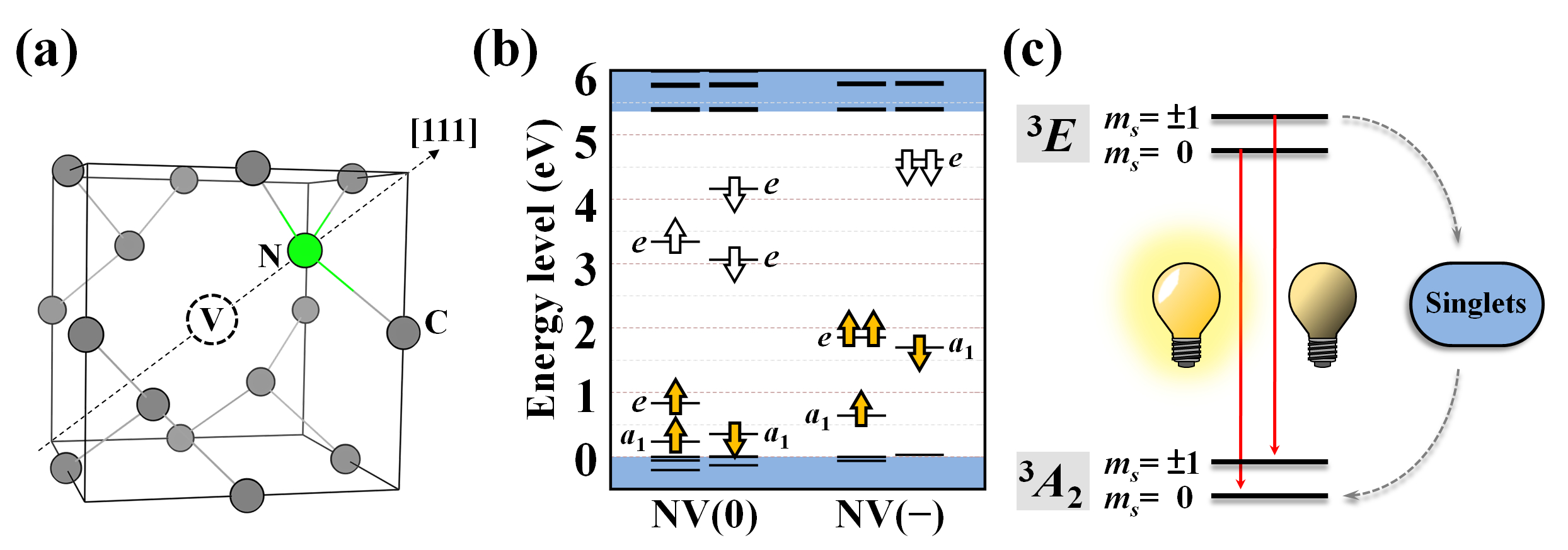}%
\caption[]{\label{fig:NVgeom}Nitrogen-vacancy (NV) defect in diamond.
(a) Optimized geometry in the core of NV center. (b) Single particle defect levels in the fundamental band gap in the ground state of NV defects. Empty(shaded) arrows depict holes(electrons).  (c) Many-body levels at room temperature and decay from the optically active  $^3E$ excited state to the $^3A_2$ groundstate of NV($-$). Radiative (non-radiative) transitions are depcited by straight (dashed-curved) arrows. Bright (dark) lamp illustrates the most (less) intense fluorescence of NV center that is the base of optical spinpolarization and readout of the $ms$ electron spin state.}
\end{figure}

Quantum bits or qubits are the building blocks of future quantum computers and nanoscale sensor devices. Special point defects with non-zero electron spin states may realize qubits in solids \cite{childress_coherent_2006, koehl_room_2011, pla_single-atom_2012} that can be well engineered by controlled implantation or irradiation techniques \cite{toyli_chip-scale_2010, falk_polytype_2013, muhonen_storing_2014, choi_depolarization_2017, choi_observation_2017}. Proximate qubits with electron-spin-electron-spin--dipole-dipole interaction may establish a quantum network in solids \cite{pfaff_unconditional_2014}. The negatively charged nitrogen-vacancy defect [NV($-$)] (see Refs.~\cite{du_preez_l_electron_1965, davies_optical_1976}) represents such a solid state qubit in diamond \cite{gruber_scanning_1997, jelezko_observation_2004, childress_coherent_2006}. The NV color center consists of a nitrogen atom substituting a carbon atom next to a vacancy [see Fig.~\ref{fig:NVgeom}(a)]. The defect forms a non-degenerate $a_1$ level and a double degenerate $e$ level in the gap that are occupied by four electrons in the negative charge state [see Fig.~\ref{fig:NVgeom}(b)], with constituting $S=1$ groundstate~\cite{Goss1996, gali_ab_2008}. The initialization and readout of this qubit can be done optically [see Fig.~\ref{fig:NVgeom}(c)], where the luminescence intensity of the illuminated NV center depends on the electron spin state, and subsequent optical cycles will polarize the electron spin to the $ms=0$ state. We note that the defect can also be found in its neutral charge state [NV($0$)] \cite{Mita1996}, in which the $e$ level is only occupied by a single electron. NV($0$) coexists with NV($-$) in diamond when the quasi Fermi-level lies around the midgap~\cite{Webber2012, DeakPRB2014}.    

Isolated NV($-$) has a long spin coherence time of $\approx1.8$~ms \cite{balasubramanian_ultralong_2009} in high quality $^{12}$C enriched diamond samples that persists at room temperature. This makes this qubit very attractive for biological or biomolecule sensing applications~\cite{haberle_nanoscale_2015, rugar_proton_2015, glenn_single-cell_2015, devience_nanoscale_2015}, where the NV centers are engineered relatively close to the surface of diamond, in order to sense the objects on the surface. A persistent problem of near-surface NV centers is the significantly reduced coherence time compared to that of deeply buried NV centers in diamond. The origin of the noise causing this effect is still under intense research~\cite{ofori-okai_spin_2012, myers_probing_2014, rosskopf_investigation_2014, romach_spectroscopy_2015, kim_decoherence_2015, Wang2016, myers_double-quantum_2017, Yamano_charge_2017}.

Recently, another configuration of NV($-$) centers has been found with reduced coherence time. In these experiments~\cite{choi_depolarization_2017, choi_observation_2017}, NV($-$) centers have been engineered with an extremely high concentration of about 45~ppm, that resulted in a strong long-range magnetic dipolar interaction of $(2\pi)$420~kHz between them. However, the observed electron spin coherence time was significantly reduced to about 67~$\mu$s~\cite{choi_depolarization_2017}. A charge fluctuation model was developed between neighboring NV($-$) and NV($0$) defects with about an average distance of 5~nm that led to the decoherence of the NV($-$) spin state. The charge fluctuation characteristic time was estimated to be $\tau=$10~ns, that corresponded well to the estimated spin depolarization frequency of $(2\pi)$3.3~MHz. The experimental facts implied tunneling-mediated charge diffusion~\cite{choi_depolarization_2017}. However, the underlying physical mechanism of charge fluctuations was not understood. Deep knowledge about the underlying physical mechanism is of high importance both for quantum network and nanosensor applications of NV center.

In this Letter, we propose a microscopic model for the source of decoherence in NV qubits, which is of high importance in quantum network as well as nanosensor applications of NV center in diamond. This model is based on the quantum mechanical tunneling or hopping of the electron of NV($-$) and a proximate acceptor defect in diamond. In order to model this charge dynamics, we consider a system composed of an NV($-$) and NV($0$) defects where NV($0$) is an acceptor, and the electron of NV($-$) can hop between the two defects. We developed an \emph{ab initio} method to calculate the corresponding hopping time $\tau$ for various configurations of NV defect pairs in diamond that resulted in an average $\tau\approx10$~ns at a distance of $\approx4.4$~nm. We estimate that a distance of $\approx9$~nm between the NV sensor and the acceptor defect is required to maintain the coherence time ($\approx1$~ms) of the isolated NV qubit in $^{12}$C enriched diamond. Our results are in good agreement with previously deduced experimental data for high concentration of NV centers \cite{choi_depolarization_2017}, and provide an explanation for one of the decoherence mechanisms of near-surface NV qubits. As an outcome of our study, we claim that proximate and isolated [NV-NV]($-$) pair can be used to study a quantum mechanical bond by optically detected magnetic resonance (ODMR) technique.

\begin{figure}[ht]
\includegraphics[width=0.45\textwidth]{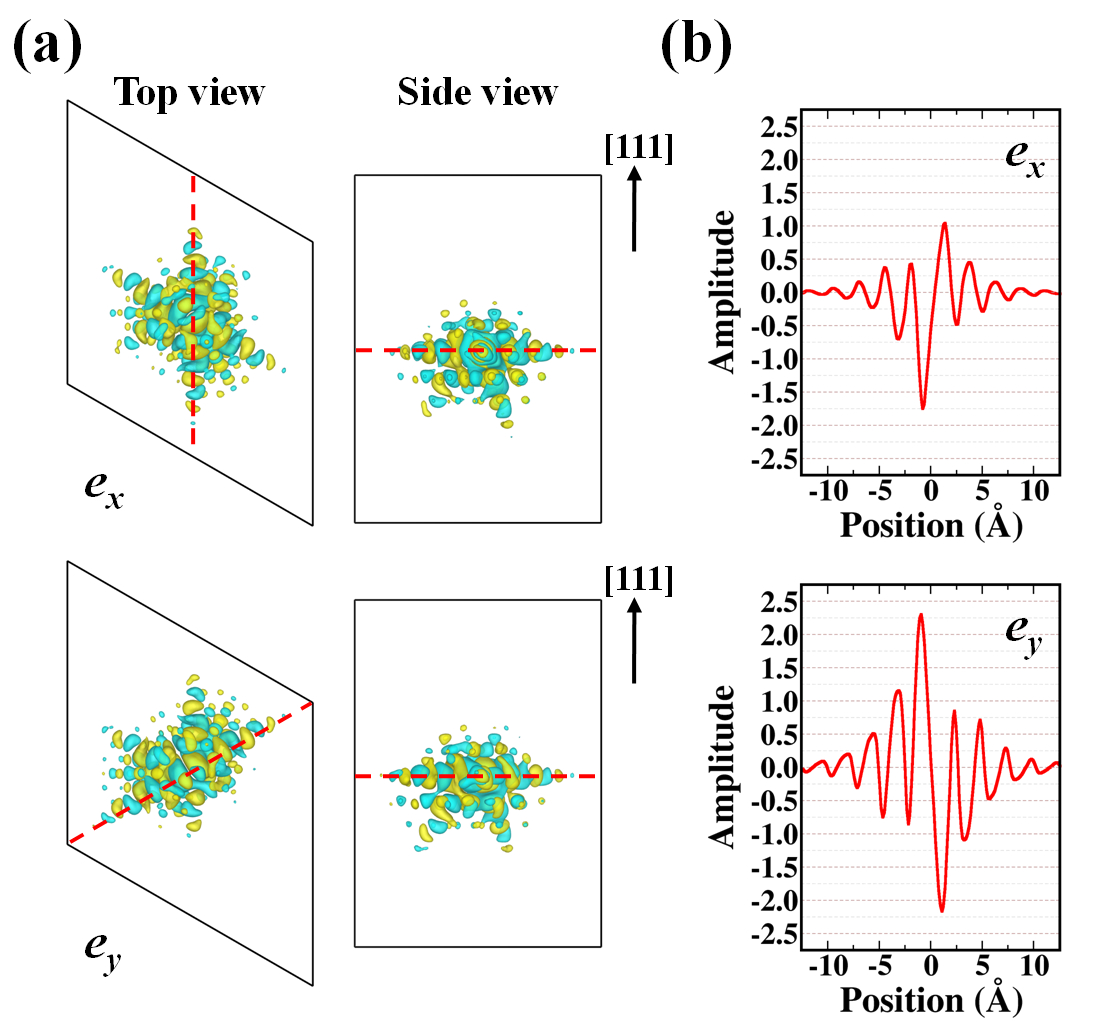}%
\caption[]{\label{fig:wfs}
Defect wavefunctions in $10\times10\times5$ hexagonal supercell of diamond with $25.1~\AA\times25.1~\AA\times30.7~\AA$ lattice constants.
(a) Top and side view of the $e_x$ and $e_y$ defect wavefunctions of NV($-$). For the sake of clarity, the atoms are not shown in diamond. The real space wavefunctions are visualized: the green and yellow lobes represent the isosurface of the wavefunction at values of 
$+1.89\times10^{-6}$~1/\AA$^3$ and $-1.89\times10^{-6}$~1/\AA$^3$ values, respectively. 
(b) The wavefunction amplitude profile along the dashed line in (a) panel. The wavefunction completely decays at the edge of the supercell boundaries. The origo is set to the middle of the dashed line.}
\end{figure}

The microscopic theory of tunneling of an electron of NV($-$) to a nearby NV($0$), is presented below. This tunneling electron of NV($-$) is originally localized on the $e$ orbital ($\psi_A$) at site $A$. NV($0$) acts as an acceptor at site $B$ in which the $e$ orbital ($\psi_B$) of NV($0$) will be occupied by this electron. The hopping rate between the two sites ($\Omega_{AB}$) may be calculated as 
\begin{equation}
\label{eq:omega}
\Omega_{AB}(r)\approx\frac{1}{\hbar}
\langle \psi_A|\hat{H}_{AB}|\psi_B\rangle_r
\approx\frac{E_0}{\hbar} \langle \psi_A |\psi_B\rangle_r
\end{equation} 
where $\hat{H}$ is an effective single particle Hamiltonian acting on the electron that binds the two sites, $r$ is the distance between the two defects, and $\hbar$ is the Planck-constant. We note that this electron will tunnel back and forth between these defects when isolated, i.e., a quantum mechanical bond is established with non-vanishing probability of the electron wavefunction between the two defects. As a consequence, the Mulliken's theory of quantum mechanical bonds~\cite{Mulliken1949} can be directly applied in the evaluation of the integral which yields an overlap integral of the real wavefunctions at $A$ and $B$ sites ($\langle \psi_A |\psi_B\rangle$) multiplied by $E_0$ where the latter is related to an average of the on-site energies of the two sites. The $\psi_{\{A,B\}}$ wavefunctions have a special shape and oscillatory decay from the $A$ and $B$ sites of the defects (see Fig.~\ref{fig:wfs}), thus the accurate evaluation of the overlap integral requires \emph{ab initio} calculations. We will show below that the value of $E_0$ can be also deduced from first principles calculations.

Our first principles approach is based on plane-wave supercell Kohn-Sham DFT\cite{hohenberg_inhomogeneous_1964,WKohn_self_1965} calculations as implemented in the \textsc{vasp} code \cite{kresse_efficient_1996} within the projector augmented wave method \cite{blochl_projector_1994}. To model the NV defect, we applied extremely large supercell with thousands of atoms that will be discussed below. The $\Gamma$-point sampling of the Brillouin-zone results in converged electron density and real Kohn-Sham wavefunctions in these large supercells. The plane wave cutoff was set to 370~eV \cite{DeakPRB2014}. We applied Perdew-Burke-Ernzerhof (PBE) DFT exchange-correlation functional \cite{perdew_generalized_1996} that provides fairly accurate wavefunctions in the groundstate of NV defects \cite{GaliPRB2009, szasz_hyperfine_2013}. We projected the values of the Kohn-Sham defect wavefunctions to the three-dimensional (3D) equidistant grid within the applied supercell with $\approx0.31$~\AA\ distances between the grid points. In the optimized geometry of the NV defects, quantum mechanical forces acting on the atoms were less than 0.01~eV/\AA .  In the calculation of $E_0$ we also used 512-atom $4\times4\times4$ cubic supercell and band structure calculations in the Brillouin-zone. 

In the calculation of the overlap integral we assume an interaction between an isolated NV defect pair. The wavefunctions of NV($-$) and NV($0$) are calculated in separate diamond clusters. Each diamond cluster is constructed from a 3000-atom hexagonal diamond supercell, which is is commensurate with the $C_{3v}$ symmetry of the defect and sufficiently large to accommodate the extent of the defect wavefunctions (see Fig.~\ref{fig:wfs}).
The line profiles in Fig.~\ref{fig:wfs}(b) show that the wavefunctions decay relatively fast, and the electron probability $|\psi|^2$ is about 0.1\% at a distance of 1~nm away from the center of the defect. We note that the same conclusion was achieved for the $e_x$ and $e_y$ defect states of NV($0$). Therefore, the calculated defect wavefunctions in this supercell and the corresponding 3D grid can be taken as a cluster of an isolated defect (see more details in the Supplemental Material~\onlinecite{supplementary}). In this construction, we could reach 5.2~nm maximum distance between the two NV defects in the (111) plane.  By taking these diamond clusters of NV($-$) and NV($0$), the corresponding overlap integral can be calculated at various distances and all the possible NV--NV orientations within the range of the given maximum distance. 
  
However, there is no strict formula for obtaining the value of $E_0$. We work around this problem by substituting the system of isolated [NV--NV]($-$) pair by the system of periodic array of NV($-1/2$) defects, where ($-1/2$) means half charge (half electron). The idea is that the periodic array of NV($-1/2$) defects constitute the same type of covalent bonds as those in isolated [NV--NV]($-$) molecule, in which the electron is equally shared between the two sites because of the same potential induced by the NV defects. These covalent bonds in the periodic array model will naturally form a defect impurity band in the diamond bandgap. A tight binding theory can be applied to the impurity band of this system, i.e., the periodic array of NV($-1/2$) embedded into diamond crystal, in which the hopping integral of the tight binding theory (see Eqs.~(6) and (7) in Ref.~\onlinecite{supplementary}) is basically identical with $\langle \psi_A|\hat{H}_{AB}|\psi_B\rangle$ in Eq.~\eqref{eq:omega}. By applying the tight binding retrofit to the dispersion relation of the impurity band, the hopping integral can be read out (see Eq.~(8) in Ref.~\onlinecite{supplementary}), and $E_0$ can be derived by dividing the values of these integrals with the overlap of the corresponding wavefunctions (see Eq.~\eqref{eq:omega}). 

\begin{figure}[h!]
\includegraphics[width=0.46\textwidth]{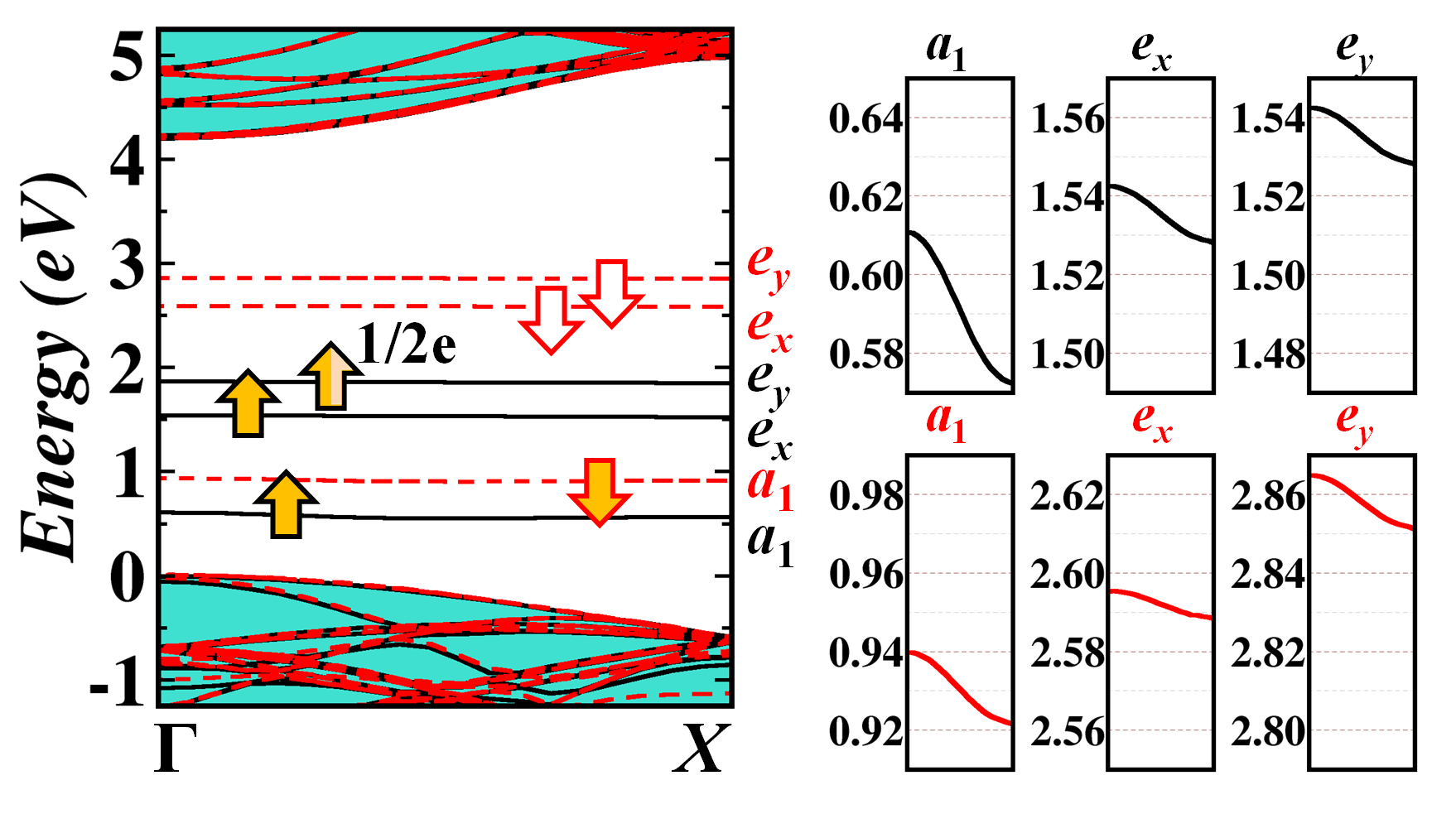}%
\caption[]{\label{fig:disp}
The band structure of NV defect in a $4\times4\times4$ cubic cell 
along $\Gamma$(0,0,0) point to $X$(1/2,0,0) point with half extra charge. The valence band maximum is aligned to zero. The calculations were carried by PBE DFT functional that provides inaccurate band gap but the Kohn-Sham wavefunctions should be accurate. Because of the half charge the symmetry is tilted a bit to $C_{1h}$ symmetry, thus $e_x$ and $e_y$ levels slightly split but it is very close to the $C_{3v}$ symmetry. The spinpolarized PBE DFT calculations results in separate spin-up (black straight curve) and spin-down (red dotted curve) bands and levels. The valence and conduction band regions are green regions. The orange and white arrows represent the occupied and unoccupied defect states, respectively. The six panels show the dispersion of each in-gap defect levels in the region $\Gamma$-$X$. The hopping integral was inferred from the dispersion of the half electron (1/2e) filled $e_y$ level.
}
\end{figure}

To this end, we choose the cubic array of NV($-1/2$) in a 512-atom cubic supercell of diamond which yields a small dispersion for the half-filled $e_y$ level (see Fig.~\ref{fig:disp}). This indicates that the direct interaction between the periodic images is very small but still sufficiently large to be above the numerical uncertainty of $\approx1$~meV. The tight binding retrofit to the $e_y$ level is almost perfect with a standard deviation of about 1\%. From these calculations we obtain the values of the hopping integrals along the corresponding directions. At these orientations and distances between the neighbor NV($-$) and NV($0$) we calculated the overlap integrals of the $e_y$ wavefunctions in the cluster model as explained previously. Finally, the ratio of the hopping integrals and overlap integrals results in 
$E_0=$~0.058~eV (see also Ref.~\onlinecite{supplementary}). This $E_0$ is used in the calculation of the hopping rates between isolated NV defect pairs as a next step.

\begin{figure}[ht]
\includegraphics[width=0.48\textwidth]{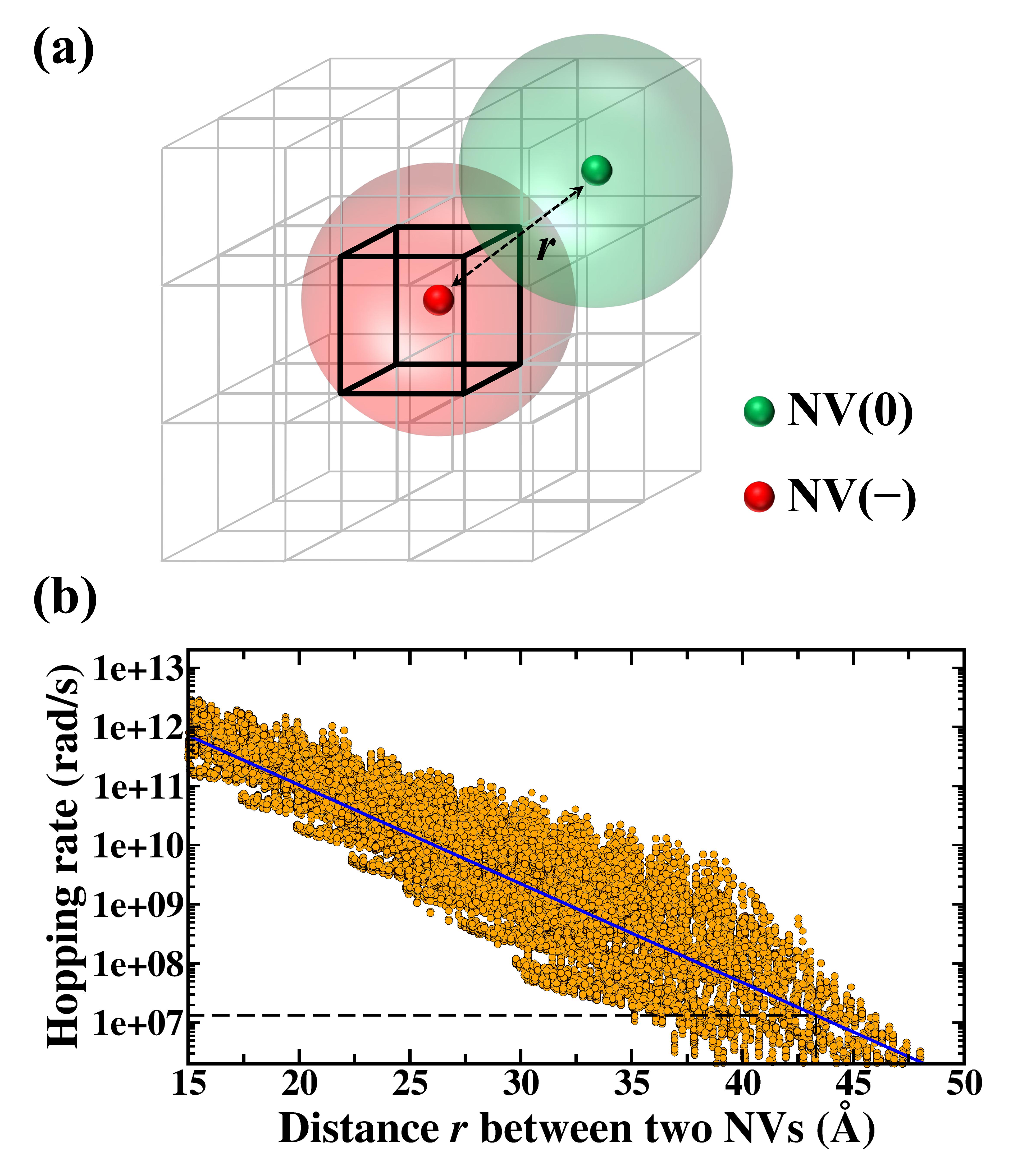}%
\caption[]{\label{fig:hopping}
(a) Schematic illustration of two NV centers with the separation of $r$. 
The wavefunction of NV is converted into a 3-dimensional equidistant grid.
The maximum of separation $r$ in the 3000-atom hexagonal cell is 5.2~nm in the (111) plane.
(b) The calculated hopping rate of electrons of all possible configurations at a given distance $r$. The average hopping rate values are shown as a blue line.
We note that all the possible configurations could be taken into account for $r \leq 4.5$~nm. For larger $r$ we have only a subset of configurations because of the constraint of the size and shape of the hexagonal cluster. The value of the hopping rate can be extrapolated at larger $r$ by assuming an overall exponential decay. The dashed horizontal line shows a critical rate at ($2\pi$)13.2~MHz  which is the deduced rate of the radiative decay (see Ref.~\onlinecite{Goldman2015}).}
\end{figure}

We computed the corresponding overlap integral at various distances (up to 5.2~nm) and all the possible NV--NV orientations when accessible in the combination of the two 3000-atom hexagonal clusters. It is apparent from Fig.~\ref{fig:wfs}(a) that the overlap of wavefunctions should be larger on the NV--NV orientation of (111) plane that that in other NV--NV orientations. Consequently, the individual hopping rates will be also larger for the corresponding NV--NV orientations. The schematic illustration of NV($-$)--NV($0$) configurations with a distance $r$ are shown in Fig.~\ref{fig:hopping}(a), and all the corresponding hopping rates are calculated and summarized in Fig.~\ref{fig:hopping}(b). By fitting the hopping rate to an exponential regression in the range from 1.5~nm to 5.2~nm, we found that the hopping rate is around ($2\pi$)13~MHz on average at 4.4~nm distance between NV defects that corresponds to $\approx10$~ns diffusion time. This is in the order of magnitude of the optical lifetime for the negatively charged NV defect in diamond~\cite{Jelezko2006}, and can explain the charge fluctuation in diamond with high NV concentration. Our calculations are in very good agreement with previous studies on such diamond samples~\cite{choi_depolarization_2017} that verifies the tunneling-mediated charge diffusion model. 

Our finding has implications in the field of quantum sensors. Our model clearly demonstrates that tunneling of the NV($-$)'s electron to a proximate defect that can accept this electron is feasible. We emphasize that this effect occurs in the ground state, without any illumination. In sensor application, the single NV qubit or NV qubit ensembles should reside near the surface of diamond for sensing the external fields. However, unwanted defects may appear on the diamond surface that can deteriorate the operation of each individual or single NV qubit. In particular, defects with acceptor levels at the surface may interact with the electron of NV center. The NV($0$) defect can serve as a model defect for acceptor-type defects at the surface that have similar extension of the acceptor wavefunctions to that of NV($0$). It was indeed shown~\cite{Kaviani2014} that such defects can exist in oxygenated diamond surfaces that have been the subject of coherence studies on near-surface NV qubits~\cite{ofori-okai_spin_2012, myers_probing_2014, rosskopf_investigation_2014, romach_spectroscopy_2015, kim_decoherence_2015, Wang2016, myers_double-quantum_2017, Yamano_charge_2017}. The derived hopping rate between NV($-$) and a nearby NV($0$) reads as ($2\pi$)$2.30\times10^{14}\exp{(-3.8 r)}$~Hz [blue line in Fig.~\ref{fig:hopping}(b)], where distance $r$ is given in nanometer unit.  Finally, we estimate that the rough critical distance between the single NV qubit sensor and the surface acceptor defect is $\approx$9~nm, in order to persist the coherence time of the qubit in the order of ms in $^{12}$C enriched diamonds. In practice, the NV sensor might be located somewhat closer than 9~nm to the diamond surface without reducing its coherence time because these acceptor defects do not dominantly reside exactly above the NV center on the surface. Here we established a tunneling-mediated model for the source of decoherence of the near-surface NV qubit.

We note that the electron will tunnel back and forth between proximate NV defects  when isolated, i.e., an [NV--NV]($-$) molecule is formed.
Consequently, isolated NV($-$)--NV($0$) pair defect can be applied to directly study quantum mechanical bonds by ODMR technique, where the tunneling rate of the electron in this system is several orders of magnitude slower (10~ns) than that of the electron in a usual molecule (femtosecond or attosecond region). By employing recent superresolution techniques~\cite{Jaskula2017}, the tunneling of the electron between the two sites can be monitored by observation of the ODMR signal of NV($-$) at both sites as a function of time. The requirement of this measurement is that the illumination applied in the ODMR measurements will not ionize NV($-$) or NV($0$).   

In conclusion, we carried out \emph{ab initio} calculations to study charge fluctuation between solid state qubits, in particular, the NV defects in diamond. We found that the electron can tunnel between proximate NV($-$) and NV($0$) defects. We provided detailed quantitative analysis on the probability of tunneling or hopping as a function of orientation and distance between the NV defects. Our findings are in quantitative agreement with data from previous experimental studies. We identified the critical distance between near-surface NV sensor and the surface defects, for maintaining the favorable coherence properties of the qubit. Our conclusions are important in future quantum network and sensing studies of this solid state qubit. Our \emph{ab initio} methodology for studying charge and spin fluctuation of proximate qubits is a template for other solid state qubit systems.  

\begin{acknowledgments}
We acknowledge the funding support from the EU Commission FP7 grant No.~611143 (DIADEMS) and the National Research Development and Innovation Office of Hungary within the Quantum Technology National Excellence Program  (Project No.~2017-1.2.1-NKP-2017-00001).    
\end{acknowledgments}


%

\end{document}